\documentclass[]{article}
\RequirePackage[OT1]{fontenc}
\RequirePackage{amsthm,amsmath}
\RequirePackage[numbers]{natbib}
\RequirePackage[colorlinks,citecolor=blue,urlcolor=blue]{hyperref}

\usepackage{lscape}
\usepackage{lineno,hyperref}
\usepackage{amsfonts}
\usepackage{graphicx,psfrag,epsf}
\usepackage{enumerate}
\usepackage{framed}
\usepackage{url} 
\usepackage{flexisym}
\usepackage{subfig}
\modulolinenumbers[5]

\usepackage{graphicx}
\usepackage{subfig}
\usepackage{amsmath}
 \usepackage{amsfonts}
\usepackage{epsfig}
\usepackage{authblk}

\usepackage{epsfig}

\newtheorem{propo}{Proposition}
\newtheorem{defi}{Definition}
\begin{document}
%
\title{Learning Gaussian Graphical Models \\by symmetric parallel regression technique}
%
\author[1]{Daniela De Canditiis}
\author[2]{ Armando Guardasole}
\affil[1]{Istituto per le Applicazioni del Calcolo ``M. Picone" - Rome,~Italy}
\affil[2]{University of ``Tor Vergata" -  Rome,~Italy}

\maketitle

\begin{abstract}
In this contribution we deal with the problem of learning an undirected graph which encodes the conditional dependence relationship between variables of a complex system, given a set of observations of this  system. This is a very central problem of modern data analysis and it comes out every time we want to investigate a deeper relationship between random variables, which is different from the classical dependence usually measured by the covariance.

In particular, in this contribution we deal with the case of Gaussian Graphical Models (GGMs) for which the system of variables has a multivariate gaussian  distribution. We study all the existing techniques for such a problem and propose a smart implementation of the symmetric parallel regression technique which turns out to be very competitive for learning sparse GGMs under high dimensional data regime.  

\end{abstract}

\medskip

\medskip
\section{Introduction}
Determining conditional independence relationships through undirected graphical models is a key component of the statistical analysis of complex observational 
data in a variety of domains such as bioinformatics, image analysis, physics, economics, etc. In many of these applications one is interested in estimating the
 undirected graphical model underlying a joint distribution of a vector of random variables which constitute a complex interacting system. In particular, in 
 this work we deal with the problem of learning a GGM (Gaussian Graphical Model) which encodes the conditional dependence relationship between variables $(X_1,\ldots,X_p) \sim \emph{N}(\mu,\Sigma)$.

It is very important to note that the conditional dependence relationship is very different from the marginal dependence relationship and that the former does not imply the second nor vice versa, as pointed out in the well know Yule-Simpson effect. More precisely, two variables $ X_i $ and $ X_j $ are conditionally independent (conditioned on the rest of the other system's variables $ X_l $ with $ l \neq i, j $) if their conditional distribution is the product of the conditional marginal distributions, while two variables are independent (in the classical sense, i.e. marginally) if their joint distribution (i.e. the marginal of $ X_i $ and $ X_j $) is the product of the marginals. The concept of conditional independence, being more sophisticated with respect to the marginal one, can capture more fundamental relations between variables and this is the reason why it is becoming central in the analysis of complex system of variables. As an example, consider a data set which consists of $p$ simultaneous protein expression levels, measured  in $n$ different cell types, hypothesizing that the joint distribution of  the $p$ proteins can be modeled as a multivariate Gaussian. Starting from the dataset, you want to discriminate between direct and indirect proteins interaction. This is a classic example of biological network, where the marginal (indirect) relationship between different proteins is almost certainly present since the system of $p$ proteins is very complex, and thus we are not interested in it; yet the relationship of conditional (direct) dependence expresses a deeper and more interesting link from the biological point of view (see \cite{Lundberg et al. 2016} for clear explanation).

In this contribution we propose an implementation of the symmetric parallel regression technique for learning a GGM, showing its performance in the case of high dimensional data. In particular, in Section 2 we formalize the problem and we describe the state of the art of the existing methods in literature. In Section 3 we study a variant of one of these methods proposing a smart algorithm for its implementation. Finally, in Section 4 we show a set of numerical tests that prove the effectiveness of the proposed algorithm.

\section{Mathematical framework and state of the art}

For a complete and exhaustive treatment of graphs theory we refer to \cite{Lauritzen}; below we give only definitions and properties necessary for this work.
A finite graph $G=(V,E)$ consists of a finite collection of \textit{nodes} $V=\{1,2...,p\}$  and a collection of \textit{edges} $E \subseteq V\times V$. For the scope of this work, we will consider graphs that are \textit{undirected}, namely graphs whose edges are not ordered, i.e. there is no distinction between the edges $(i,j)$ and $(j,i) \in E$. Moreover, for any $i \in V$ $N(i) := \{j \in V : (i,j) \in E\}$ is the set of neighbours of node $i$ and $C \subset V$ is a \textit{clique} if $(i,j)\in E$ for all $i,j \in C$ such that $i \neq j$.  

In this paper the notion of a graph is used to keep track of the conditional dependence relationship between random variables of a complex system.
By complex system here we mean a jointly distributed  vector of random variables $(X_1,X_2,...,X_p)$ that interact with each other. Moreover a formal definition of conditional independence relationship is  the following:

\begin{defi}\label{def1}  Two random variables ($X_i$, $X_j$) of a random vector $(X_1, X_2, \ldots,X_p)$ are conditionally independent, $ X_i  \perp X_j | X_{V\backslash\{i,j\}}$, if

\begin{equation}
\begin{array}{c}
\quad f(X_i,X_j | X_{V\backslash\{i,j\}})=f(X_i | X_{V\backslash\{i,j\}}) f(X_j | X_{V\backslash\{i,j\}}) \\
\Updownarrow \\
f(X_i | X_{V\backslash\{i\}}) \mbox{   does not depend from   } X_j
\end{array}
\end{equation}

where $f(\cdot)$ stands for density distribution  and $X_{S}:=(X_s, s \in S)$.
\end{defi}
 Associated with an undirected graph $G=(V,E)$ and a system of random variables $X_V$ indexed in the vertex set $V$ there is a range of different Markov properties which establish how much the graph is explanatory of the conditional independence property of the random variables, see \cite{Lauritzen} for details. Specifically, in this work we deal only with system of random variables which are global Markov with respect to an undirected graph $G=(V,E)$, and in particular it holds that

$$
X_i  \perp X_j | X_{V\backslash\{i,j\}} \quad \Leftrightarrow \quad (i,j) \notin E,
$$  

which establish conditional independence among two variables $X_i$ and $X_j$ iff their corresponding nodes in the graph $G$ are not connected, as well as the fact that any variable of the system is conditional independent from the set of variables indexed in $V\backslash\{i\} \cup N(i)$ given the set of variables indexed in $N(i)$, ie
$
X_i  \perp X_{V\backslash\{i\} \cup N(i)} | X_{ N(i)}.
$

Our perspective is inferential, therefore, given a statistical sample extracted from the unknown distribution $ f (X_1,X_2,\ldots,X_p) $, we want to learn as much as possible about it. The density estimation problem is really impossible in high dimension ($p>4$) unless you make very strong assumptions, and therefore in large dimensions you are content to learn the dependence/independence conditional relations between the system variables. Learning these relationships means learning the structure of the graph for which the distribution is global Markov, but even this problem turns out to be very difficult unless you put yourself in one of the following two hypotheses: i) the distribution of the system of variable is Gaussian, ii) the distribution of the system of variable is finite discrete with non zero probability mass function.

In this paper we deal with the first case, so our working hypothesis is that $(X_1, \ldots,X_p) ~ \sim ~ N(0,\Sigma)$. We stress that the zero-mean hypothesis is not restrictive at all because we always can center data before starting analysis; moreover from now on we also suppose $\sigma_{ii}=1$ i.e. the variables are considered standardized so covariance $\sigma_{ij}$ between two variables is indeed correlation. Again this is not a restrictive hypothesis, because we can  standardize the columns of any data matrix before starting analysis.

\subsection{Gaussian Graphical Models}
Before to deal with the inference aspect, we recall some population results for the GGMs.
Suppose $X \sim  N(0,\Sigma)$ with $\Sigma$ strictly positive definite, then we can write its distribution  in terms of parameter $\Sigma^{-1}$ as classically :
$$
f(X)=\frac{1}{(2 \pi)^{p/2} \det(\Sigma)^{1/2}} \exp\left(-\frac{1}{2} x^t \Sigma^{-1} x\right),
$$ 
or equivalently in terms of $\Theta=\Sigma^{-1}$
\begin{equation} \label{eq:factorization}
f(X)=\left( \frac{\det(\Theta)}{(2 \pi)^p}\right)^{1/2} \exp \left(-\frac{1}{2} \sum_{i,j=1}^p \theta_{ij} x_i x_j\right).
\end{equation}
From the remarkable Hammersley-Clifford theorem, it follows that, being $f(X)>0$, the \emph{global Markov property} with respect to an undirected graph $G=(V,E)$ is equivalent to the \emph{factorization property} over $G=(V,E)$, i.e. 
\begin{equation}\label{eq:factorization_general}
f(X)=f(X_1,\ldots,X_p)= \frac{1}{Z} \prod_{C \in \mbox{\emph{C}}} \psi_C(X_C)
\end{equation}
where \emph{C} is the set of all possible cliques of the graph $G$ and $\psi_C(X_C)$ is a real-valued function of the subvector $X_C:=(X_s, s \in C)$ taking positive values. Then from eq.(\ref{eq:factorization}) it follows that $f(X)$ factorizes as a product of strictly positive and real-valued functions, so the knowledge of the support of $\Theta$ is equivalent to the knowledge of $G=(V,E)$ with respect to which the distribution $f(X)$ is global Markov. 
This is a very important fact, because it allows to assert that two variables $X_i$ and $X_j$ are conditional independent, i.e. nodes $i$ and $j$ are not connected into graph $G=(V,E)$ if and only if $\Theta_{ij}=\Theta_{ji}=0$. This fact can also be derived directly by the property of multivariate Gaussian distribution, as claimed in the following proposition:
\begin{propo} \label{prop1}
If $X=(X_1,\ldots,X_p)  \sim N(0,\Theta^{-1})$, then for any $j \in \{1,2,...,p\}$, the  distribution of  $X_j$ given the rest is still Gaussian with mean and variance given by 
$$
E(X_j | X_{V\backslash\{j\}})=- \sum_{i \neq j} \frac{\Theta_{ij}}{\Theta_{jj}} X_i \quad \mbox{and} \quad var( X_j | X_{V\backslash\{j\}})= \Theta_{jj}^{-1}.
$$
\end{propo}
The proof can be obtained in Lemma A.4 of page 215 of \cite{Giraud}.

From Proposition \ref{prop1} and Definition \ref{def1} it follows that $X_j$ is conditional independent from $X_i$  iff $\Theta_{ij}=0$. 

We can now turn to the inferential aspect we are interested in. 
Suppose we have a random sample from a $N(0,\Sigma)$, i.e. suppose we have a data matrix $\mathcal{X}$ of dimension $n \times p$ where each row is a realization of this random variable. The objective of our analysis is to estimate the graph $G=(V,E)$ for which the unknown distribution is global Markov. For previous results, we can equivalently state our problem as the following:

Given $\mathcal{X}$, estimate the support of $\Theta=\Sigma^{-1}$. In the following sections we present the most used methods to solve this problem together with a variant of one of these that turns out to be more advantageous not only from the performance point of view but especially from the computational point of view. 

\subsection{Estimating G by multiple testing} 

The simplest method to estimate the support of $\Theta$ is to invert numerically an estimate of $\Sigma$ and then test if its coefficients are zero. As a first  step, given the data matrix $\mathcal{X}$, we have to  evaluate estimator $\hat{\Sigma}= \mathcal{X}^t \mathcal{X} / n$. Since the data are standardized, we observe that $\hat{\Sigma}$ is indeed an estimator for the correlation matrix, then $\hat{\Sigma}^{-1}$ is indeed proportional to an estimator of the conditional correlation. Denote $[ \star]_{ij}$ the $ij$-th entry  of matrix $\star$, then $\hat{\rho}_{ij}=-[\hat{\Sigma}^{-1}]_{ij} / \sqrt{ [\hat{\Sigma}^{-1}]_{ii} [\hat{\Sigma}^{-1}]_{jj}}$ is an estimate of the conditional correlation between variables $i$ and $j$. When  $cor(X_i,X_j | X_{V\backslash\{i,j\}})=0$, we have (see \cite{Anderson}, Chapter 4.3)
\begin{equation} \label{eq:student}
\hat{t}_{ij}= \sqrt{n-p-2} ~ \frac{\hat{\rho}_{ij}}{\sqrt{1-\hat{\rho}_{ij}^2}}  \sim Student (n-p-2);
\end{equation}

then, for each $i \neq j$ we can test the hypothesis 
$$
H_0: ~ cor(X_i,X_j | X_{V\backslash\{i,j\}})=0 \leftrightarrow (i,j) \notin E 
$$ 
 by using the test statistic in eq. (\ref{eq:student}). It is instructive to observe that, for Gaussian variables, independence is equivalent to zero correlation and this is true also for conditional distribution which are still Gaussian as claimed in Proposition \ref{prop1}.
 
While the empirical variance (in this case correlation) estimator $\hat{\Sigma}$ does not suffer of instability when the dimension $p$ gets larger, its inversion become more and more unstable, being not invertible at all in the case $p>n$. Hence alternative  approaches have been proposed to deal with the GGMs learning problem in the high dimensional case and they are the object of the following sections.

\subsection{Estimating G by maximum likelihood penalized technique} \label{sec:GL}
Try to infer graph $G$  is hopeless in the high dimensional setting without additional structural assumption, hence from now on we suppose that the underlying graph is sparse (it has a few edges). Since $\Theta_{ij}=\Sigma^{-1}_{ij}=0$ when there is no edge between \textit{nodes} $i$ and $j$, the sparsity of $G$ translates into coordinate sparsity for matrix $\Theta$. 
Given $\mathcal{X}$ whose rows represent $n$ samples from a zero-mean multivariate Gaussian distribution with $\Theta=\Sigma^{-1}$, we can write the Log-likelihood function using expression in eq. (\ref{eq:factorization}) and standard property of the trace operator
\begin{equation} \label{eq:Likelihood}
\mbox{\textsc{L}}(\Theta;\mathcal{X})= \frac{1}{n}\sum_{i=1}^n \log(f(\mathcal{X}_{i \cdot})) ~ \propto ~ \log(\det(\Theta)) - tr(\hat{\Sigma} \Theta),
\end{equation}

where $\hat{\Sigma}=\mathcal{X} \mathcal{X}^t /n$ is the empirical covariance matrix. The standard theory of $MLE$ (Maximum Likelihood Estimator) suggests to maximize function in (\ref{eq:Likelihood}), however since we are seeking for GGMs based on sparse graphs, in order to control the number of non-zeros entry of the MLE of matrix $\Theta$ the following $\mbox{\textit{l}}_1$-penalization approach is considered 
\begin{equation} \label{eq:grahicalLasso}
\hat{\Theta} =  \mbox{argmax}_{\Theta}  \left\{ \log(\det(\Theta)) - tr(\hat{\Sigma} \Theta) - \lambda \sum_{i \neq j} |\Theta_{ij}| \right\}.
\end{equation}   
We point out that diagonal elements $\Theta_{ii}$ are not penalized because they are not expected to be zero. Solution of (\ref{eq:grahicalLasso}) has been studied by many authors but only in \cite{Friedman et al} a smart first-order block coordinate-descendent algorithm has been proposed that made this technique famous with the name of \textit{Graphical Lasso} or \textit{glasso}. 
 
\subsection{Estimating G by parallel regression technique}   

Although the  algorithm proposed in \cite{Friedman et al} is efficient, in high-dimensional regime it can be less competitive; therefore in \cite{Bulmann} an alternative approach for learning GGMs has been proposed under sparsity hypothesis.

Let us first observe that from Proposition \ref{prop1}, for each $j \in \{1,\ldots,p\}$, there exists $\epsilon_j \sim N(0, \Theta_{jj}^{-1})$ independent of $\{X_i : i \neq j\}$, such that $X_j= - \sum_{i \neq j} \frac{\Theta_{ij}}{\Theta_{jj}} X_i +\epsilon_j$. Denote $\beta_{ij}=-\Theta_{ij} / \Theta_{jj}$ with $i \neq j$, hence an estimate of $\beta_{\cdot j}$ can be obtained as the LS (Least Square) solution of the classical regression problem
\begin{equation} \label{eq:oneregression}
\hat{\beta}_{\cdot j}=  \mbox{ argmax}_{\beta \in R^{p-1}}\frac{1}{2n} \| \mathcal{X}_{j} - \mathcal{X}_{V\backslash\{j\}} \beta\|_2^2
\end{equation}
where $\mathcal{X}_S$ is the sub matrix of $\mathcal{X}$ with columns indexed in $S$. Since $\beta_{ij}$ is a scalar multiple of $\Theta_{ij}$, if $\beta_{ij}=0$  variables $X_i$ and $X_j$  are conditional independent, i.e. there is no edge between nodes $i$ and $j$; hence authors in \cite{Bulmann} propose to learn $N(i)=\{ j \neq i: (i,j) \in E \}$ adding a $\mbox{\textit{l}}_1$-penalty term in criterion(\ref{eq:oneregression}) to enforce sparsity. Formally for each $j=1,\dots,p$ the authors solve
\begin{equation} \label{eq:oneregressionlasso}
\hat{\beta}_{\cdot j}= \mbox{argmax}_{\beta \in R^{p-1}}\frac{1}{2n} \| \mathcal{X}_{j} - \mathcal{X}_{V\backslash\{j\}} \beta\|_2^2 + \lambda \| \beta \|_1.
\end{equation}
Unfortunately, there is a difficultly in order to learn $G$ from such an approach, because there is no constrain enforcing that $\hat{\beta}_{ij}=0$ when $\hat{\beta}_{ji}=0$, hence it is possible that node $j$ is a neighbour of node $i$ and not vice versa. So we have to choose an arbitrary decision rule in order to construct $\hat{E}$ an estimate of the edges set, for example in this paper we adopt the rule $(i,j) \in \hat{E}$ iff $\hat{\beta}_{ij} \neq 0$ OR  $\hat{\beta}_{ji} \neq 0$.
 
This method gives very good results and it is much less computational expensive with respect to \textit{glasso}. Its efficiency is due especially to the fact that it is a node-wise approach learning the neighbours of each node separately, while  \textit{glasso} is a global approach learning the whole graph. Finally, it is important to stress that this parallel regression method can be reformulated in term of a unique multivariate regression problem. More precisely, denote $\mathcal{B}$ the space of $p \times p$ matrices with zero diagonal and $\hat{B}$ the zero diagonal matrix whose $j$-th column has extra-diagonal elements equal to  $\hat{\beta}_{\cdot j}$ defined in (\ref{eq:oneregressionlasso}), then the $p$  regression problems can be expressed in a unique multivariate regression problem as:
\begin{equation} \label{eq:nodewise}
\hat{\Theta}={\mbox{argmin}}_{B \in \mathcal{B}} \left\{ \frac{1}{2n} \|\mathcal{X}-\mathcal{X} B \|^2_F + \lambda \sum_{i \neq j} | B_{ij}| \right\}.
\end{equation}

\section{Estimating G by symmetric parallel regression technique}

Looking at model (\ref{eq:nodewise}) we can immediately see that it is separable, that is, the $p$ parallel regressions are in fact independent of each other. It is clear, however, that from an information point of view the $p$ regressions are not unrelated to each other because the conditional independence relationship is symmetric and therefore if the variable $X_j$ is zeroed in the regression on $X_i$ we expect that the variable $X_i$ is zeroed in the regression on $X_j$. This information can be included into the estimation procedure replacing the $\mbox{\textit{l}}_1$-penalty by a grouped penalty as proposed in \cite{report}. Hence, in this contribution we study the following variant of the parallel regression technique:
\begin{equation} \label{eq:nodewise_grouped}
\hat{\Theta}={\mbox{argmin}}_{B \in \mathcal{B}} \left\{ \frac{1}{2n} \|\mathcal{X}-\mathcal{X} B \|^2_F + \sqrt{2}\lambda \sum_{i < j} \sqrt{ B_{ij}^2 +B_{ji}^2} \right\}.
\end{equation}
Note that $\sqrt{2}$ takes into account the group size. This estimator has the nice property to be coordinate sparse with symmetric zeros, hence it is clear why we call it the symmetric parallel regression technique.
The minimization problem (\ref{eq:nodewise_grouped}) is convex, but it cannot be split in $p$ parallel subproblems, hence it is  computationally more intensive. However, in this contribution we implement a block-wise descending algorithm inspired by the general algorithm presented in \cite{Breny} which  turns out to be very interesting for learning GGMs.

\subsection{Algorithm} 

In this section we describe the algorithm obtained by adapting the general methodology presented in \cite{Breny}.
We fix $\lambda$, and consider $\hat{\Theta}$ defined in (\ref{eq:nodewise_grouped}). As already mentioned, we assume data matrix $\mathcal{X}$ standardized, i.e. 
$\sum_{i=1}^n \mathcal{X}_{i j}/n=0$ and $\mathcal{X}_{\cdot j}^t \mathcal{X}_{\cdot j}/n=1$ for each $j=1,\ldots,p$.
The general methodology proposed in \cite{Breny} provides for the updating of a group of variables at a time in a cyclical fashion until convergence is achieved.
In our case, each group of variables consists of a symmetric pair of matrix B, example $(B_{ab}, B_{ba})$ with $a <b$, and therefore the total number of groups is $ p (p-1) / 2 $. The reasoning behind this strategy is that the problem (\ref{eq:nodewise_grouped}) can be separated into $ p (p-1) / 2 $ subproblems, each of which has only two variables and can therefore be easily solved by thinking of all the others frozen in the previous step. 
For our convenience,  rewrite criterion (\ref{eq:nodewise_grouped}) in the following form:
\begin{equation} \label{eq:nodewise_grouped2}
\hat{\Theta}={\mbox{argmin}}_{B \in \mathcal{B}} \left\{ \frac{1}{2n} \sum_{j=1}^p \|\mathcal{X}_{\cdot j}-\sum_{k \neq j} B_{kj} \mathcal{X}_{\cdot k} \|^2_2 + \sqrt{2}\lambda \sum_{i < j} \sqrt{ B_{ij}^2 +B_{ji}^2} \right\}
\end{equation}
For example, let us minimize (\ref{eq:nodewise_grouped2}) in the variable $(B_{ab}, B_{ba})$. When $\sqrt{B_{ab}^2 +B_{ba}^2} \neq 0$ we can evaluate  the following partial gradient $\nabla_{ab}$ of criterion (\ref{eq:nodewise_grouped2}) with respect to the variables $(B_{ab}, B_{ba})$:
$$
\nabla_{ab}= -\frac{1}{n} \left( \begin{array}{c}    
\mathcal{X}_{\cdot a}^t (\mathcal{X}_{\cdot b} - \sum_{k \neq b} B_{kb} \mathcal{X}_{\cdot k}) \\
\mathcal{X}_{\cdot b}^t (\mathcal{X}_{\cdot a} - \sum_{k \neq a} B_{ka} \mathcal{X}_{\cdot k})
\end{array}  \right) + \frac{\sqrt{2}\lambda}{\sqrt{ B_{ij}^2 +B_{ji}^2}} \left( \begin{array}{c}
B_{ab} \\ B_{ba}
\end{array} \right).
$$

Define $z=\left( \begin{array}{c}
z_{ab} \\ z_{ba}
\end{array} \right)$ with 
$$
z_{ab}=\frac{1}{n} \mathcal{X}_{\cdot a}^t (\mathcal{X}_{\cdot b} - \sum_{k \neq a,b} B_{kb} \mathcal{X}_{\cdot k}) \quad \mbox{and} \quad z_{ba}=\frac{1}{n} \mathcal{X}_{\cdot b}^t (\mathcal{X}_{\cdot a} - \sum_{k \neq a,b} B_{ka} \mathcal{X}_{\cdot k}),
$$
hence minimizing criterion (\ref{eq:nodewise_grouped2}) in the variables $(B_{ab}, B_{ba})$ gives
\begin{equation} \label{eq:multivariateSoft}
\left( \begin{array}{c}
\hat{B}_{ab} \\ \hat{B}_{ba}
\end{array} \right) = \left(1-\frac{\sqrt{2} \lambda}{\| z\|}  \right)_+ \left( \begin{array}{c}
z_{ab} \\ z_{ba}
\end{array} \right).
\end{equation}
Solution (\ref{eq:multivariateSoft}) is known as multivariate (here $2$-variate) Soft Threshold. In conclusion, the algorithm repeats the step just described for each pair of variables in a cyclic fashion until convergence is achieved. In this contribution the convergence is achieved if a maximum number of iteration steps is exceeded or if the norm of the difference between the current $B$ and that calculated in the previous step is smaller than a certain threshold.

\section{Numerical experiments}

In this section we show the performance of the methods discussed in terms of edge reconstruction and computational time. We focus on high-dimensional regimes where method of Section 2.2 can not be applied being the empirical covariance matrix not invertible. So we will focus on the following three methods \textit{Graphical Lasso} presented in Section 2.3, \textit{Parallel Regression} presented in Section2.4 and \textit{Symmetric Parallel Regression} presented in Section 3, here denoted $GL$, $PR$ and $SPR$ respectively.
For all methods the choice of $ \lambda $ is crucial and it can make the difference, so in order to be fair in our comparative study we fix $\lambda = \log (p) / n $ which is known from the theory to be order of the optimal parameter.
Since the goal of the methods is to correctly identify the undirected graph which encodes the conditional independence relations among variables, 
i.e. to correctly identify the support of matrix $\Theta=\Sigma^{-1}$, we measure the  method performance by the following index:
\begin{equation} \label{eq:accuracy}
accuracy= (TP + TN) / (TP + TN + FN + FP),
\end{equation}
where $TP$ is the number of edges present in the graph and correctly identified (i.e. $\Theta_{ij} \neq 0 \wedge \hat{\Theta}_{ij} \neq 0$),
$TN$ is the number of edges not present in the graph and correctly identified  (i.e. $\Theta_{ij} = 0 \wedge  \hat{\Theta}_{ij}=0$),
$FN$ is the number of edges present in the graph and not correctly identified (i.e. $\Theta_{ij} \neq 0 \wedge  \hat{\Theta}_{ij} =0$) and
$FP$ is the number of edges not present in the graph and not correctly identified  (i.e. $\Theta_{ij} =0\wedge  \hat{\Theta}_{ij} \neq 0$). In all the previous definitions $\hat{\Theta}$ is the estimator obtained in eqs (\ref{eq:grahicalLasso}), (\ref{eq:nodewise}) and (\ref{eq:nodewise_grouped2}) respectively.
Note that measure in (\ref{eq:accuracy}) is a scaled measure inherit from the binary classification literature, $0 \leq accuracy \leq 1$, being more accurate methods with higher accuracy.

Aim of this section is to show how the $SPR$ technique can be competitive with the others two methods in high dimensional problems especially from a computational point of view. For that reason we analyse two different high dimensional scenarios: \textit{not severe} and \textit{severe} regime. If $p$ is the problem dimension and $n$ is the number of data, for \textit{not severe} regime we intend $n \sim p$, while for \textit{severe}  regime we intend $n << p$.
In particular in this section we analyse the following two situations $p=32$ with $n=32$ and $n=16$ respectively.
We have conducted experiments for many types of graphs and here we report results for three  graphs representing different type of categories, more precisely we show results for the following three graphs:
   
 \begin{description}
\item[G1] A \textit{Chain graph} where each node has degree 2: $\Theta_{i,i}=1$, $\Theta_{i-1,i}=\Theta_{i,i-1}=0.2$. (see Fig.1)
\item[G2] A \textit{Grid graph } where each node has degree at most 4.(see Fig.2 left)
\item[G3] A \textit{Star graph} where there is an hub node with maximum degree: $\Theta_{i,i}=1$, for all $i$ and $\Theta_{1,j}=\Theta_{j,1}=0.1$  for $j \neq 1$. (see Fig.2 right)
\end{description}
\begin{figure} \label{fig_G1}
     \center{\begin{tabular}{c}
     \includegraphics[width=8cm, height=2cm]{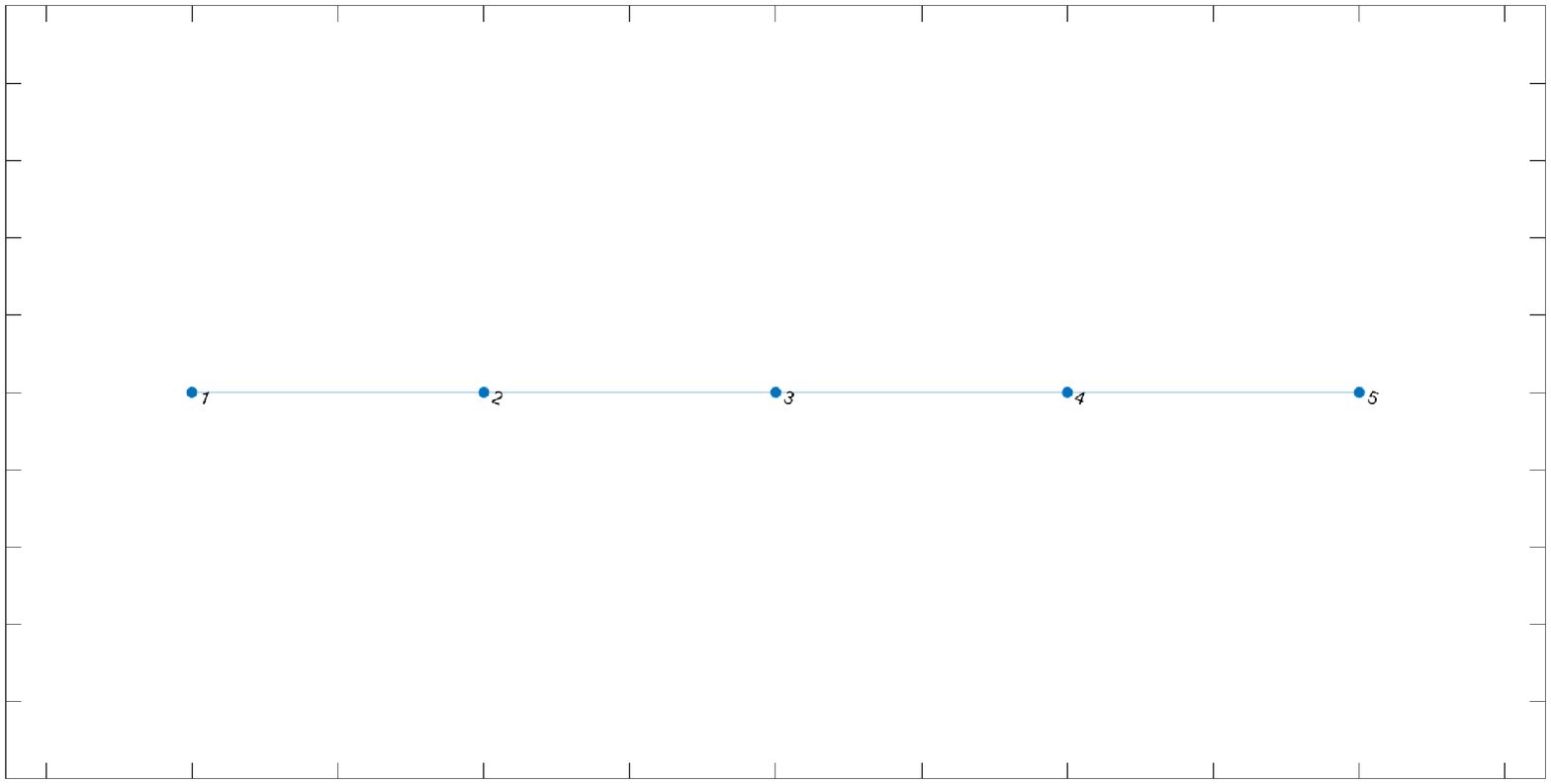} 
     \end{tabular}}
     \caption{$\mathbf{G1}$: Chain graph.}
\end{figure}
\begin{figure} \label{fig_G23}
     \center{\begin{tabular}{c}
     \includegraphics[width=10cm, height=5cm]{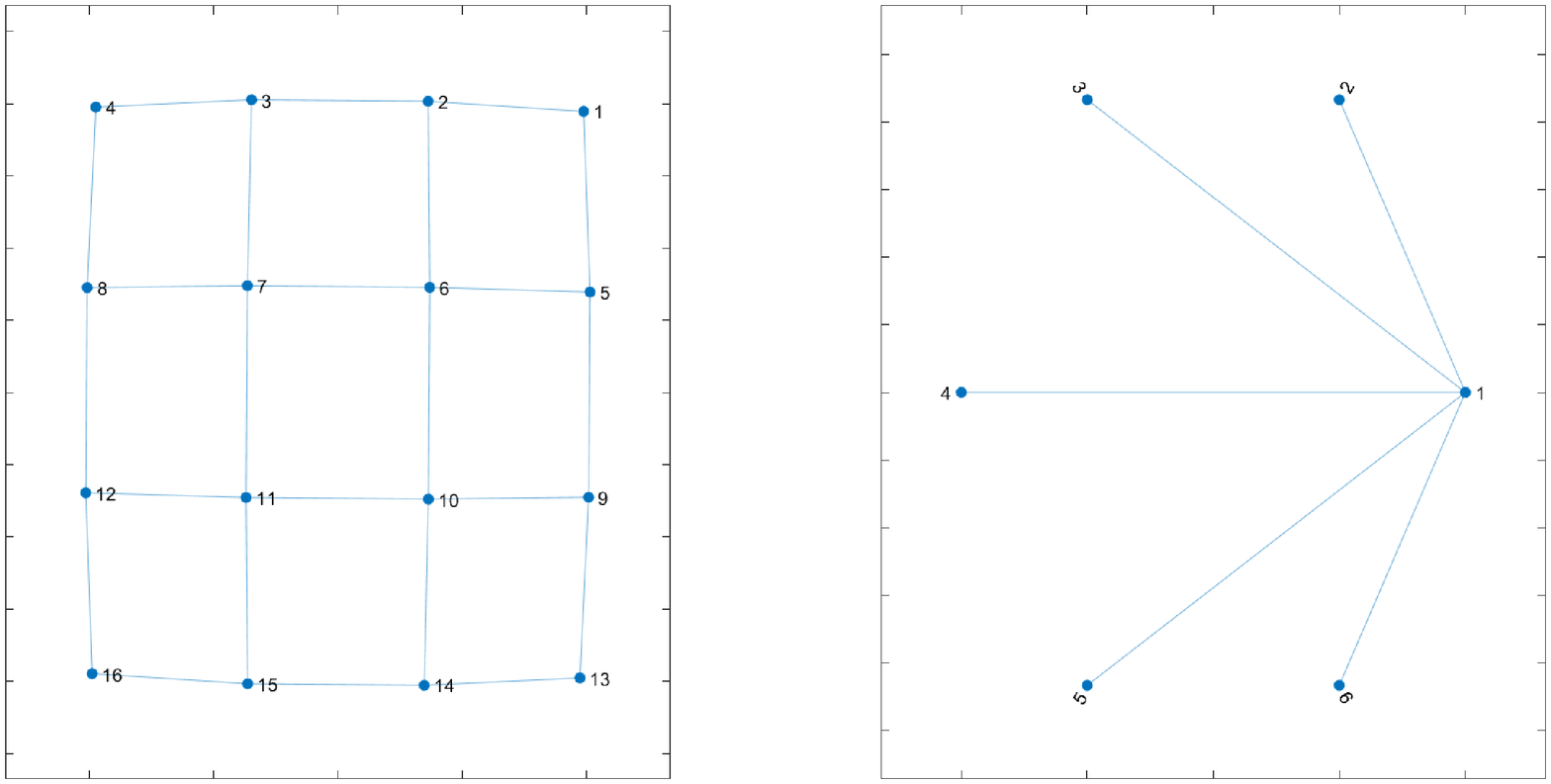} 
     \end{tabular}}
     \caption{$\mathbf{G2}$: Grid graph (left) and $\mathbf{G3}$: Star graph (right).}
\end{figure}
In figure 3, 4 and 5 we report results in terms of performance and computational time for graph $G1$, $G2$ and $G3$ respectively. First of all we note a certain robustness of results among the three different types of graph, secondly in terms of performance we note that method $SPR$ is not highly competitive with respect to the others, however it is clear how its implementation is really competitive with respect to the others methods. Finally, it is worth noting that, the $GL$ method is more expensive because it also offers a good estimate of matrix $ \Theta $ and not just of its support.
Then, at least for some situations,  we can conclude that the $SPR$ method can be competitive with the existing methods for learning the structure of a GGM under sparsity hypothesis and high dimensional regime.

The Matlab codes used to produce results of this contribution are available at http://www.iac.cnr.it/~danielad/software.html.
\begin{figure} \label{fig_chain}
     \center{\begin{tabular}{c}
     \includegraphics[width=10cm, height=7cm]{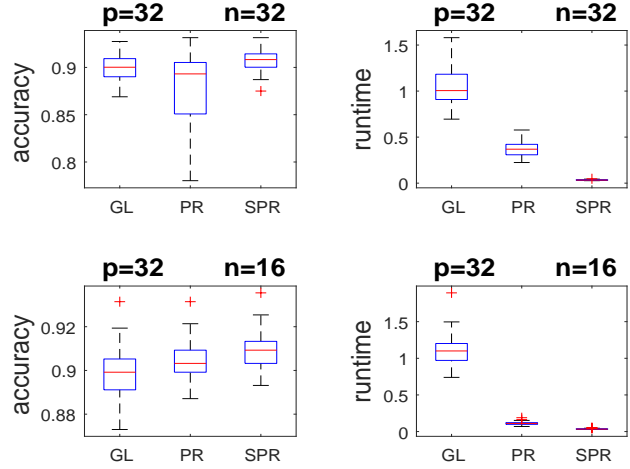} 
     \end{tabular}}
     \caption{Boxplot for performance (left side) and runtime (right side) in not severe (top) and severe (bottom) high dimensional regime. The true Graph is $\mathbf{G1}$. Results are obtained using 20 different independent data set.}
\end{figure}
\begin{figure} \label{fig_d4}
     \center{\begin{tabular}{c}
     \includegraphics[width=10cm, height=7cm]{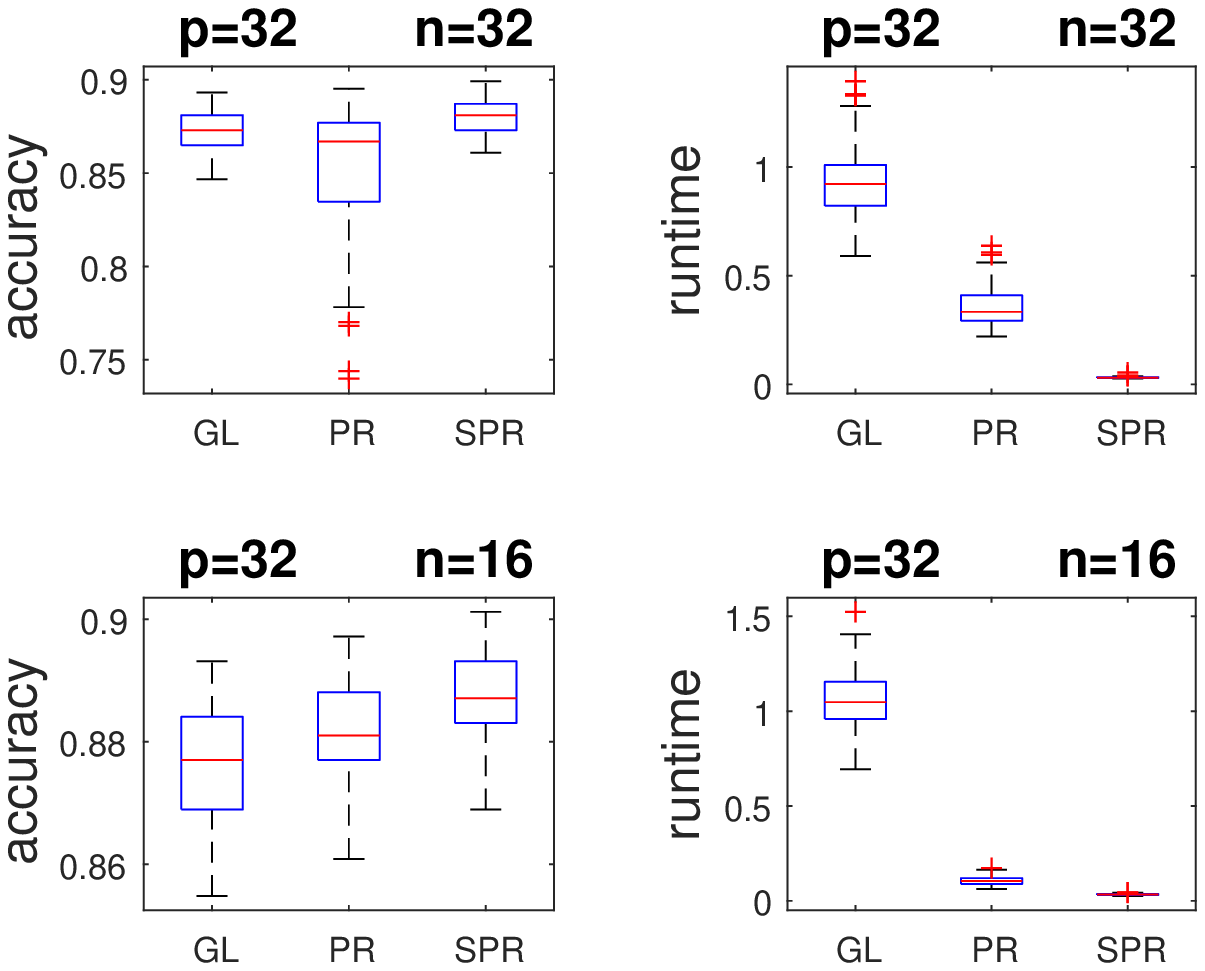} 
     \end{tabular}}
 \caption{Boxplot for performance (left side) and runtime (right side) in not severe (top) and severe (bottom) high dimensional regime. The true Graph is $\mathbf{G2}$. Results are obtained using 20 different independent data set.}
\end{figure}
\begin{figure} \label{fig_star}
     \center{\begin{tabular}{c}
     \includegraphics[width=10cm, height=7cm]{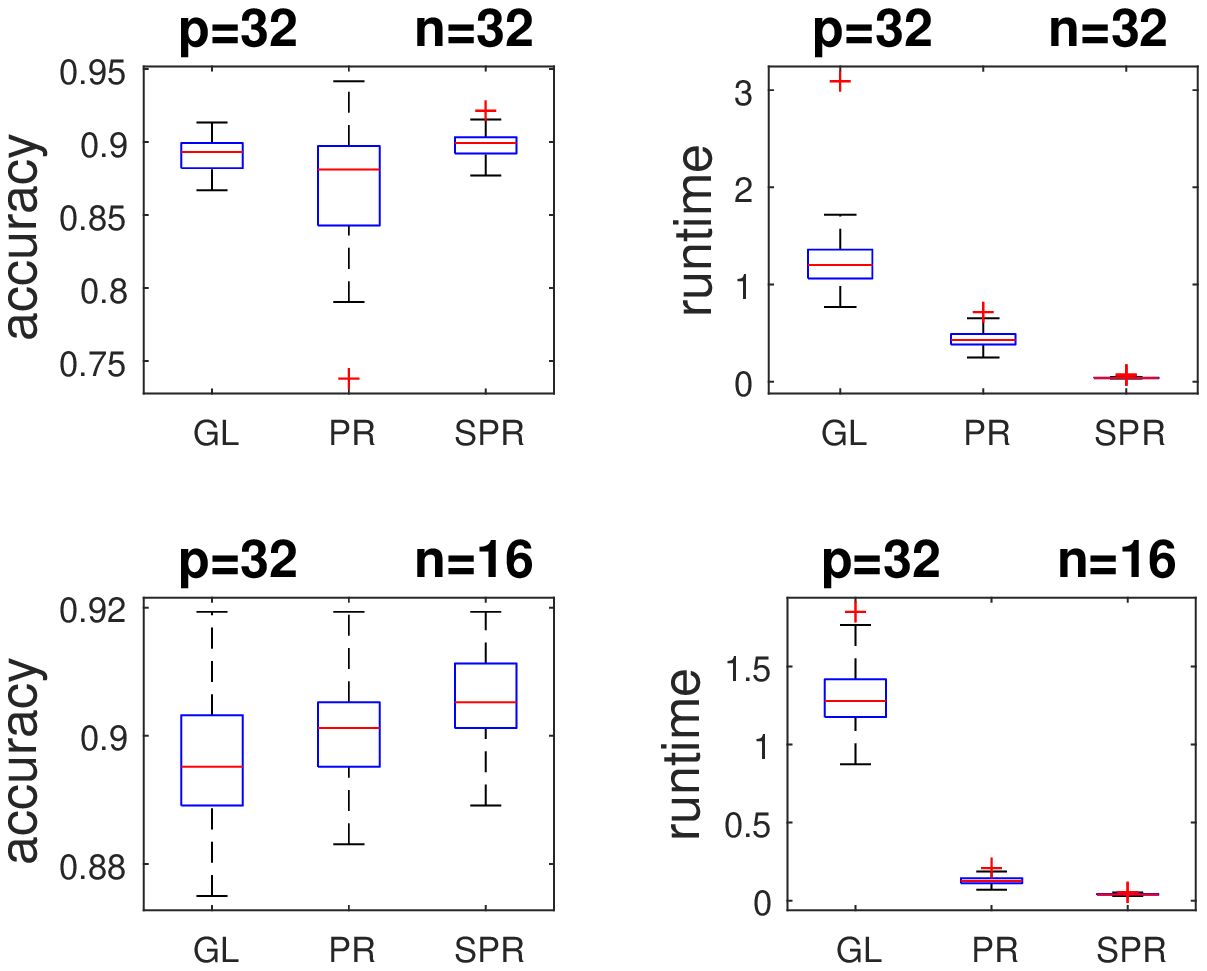} 
     \end{tabular}}
 \caption{Boxplot for performance (left side) and runtime (right side) in not severe (top) and severe (bottom) high dimensional regime. The true Graph is $\mathbf{G3}$. Results are obtained using 20 different independent data set.}
\end{figure}

\end{document}